\documentclass[prl,aps,showpacs,twocolumn,superscriptaddress,notitlepage]{revtex4}
\usepackage{amsmath}
\usepackage{graphicx}
\usepackage{bm}
\usepackage{amsbsy}
\usepackage{amsfonts}
\usepackage{amsthm}
\usepackage[dvips]{color}
\usepackage{epsfig}
\usepackage{graphicx}
\usepackage{bm}%bold math

\begin{document}

\title{Proposal to demonstrate the non-locality of Bohmian mechanics with entangled photons
}

\author{Boris Braverman}
\address{Department of Physics, MIT-Harvard Center for Ultracold Atoms, and Research Laboratory of Electronics, Massachusetts Institute of Technology, Cambridge, Massachusetts 02139, USA}
\address{Institute for Quantum Information Science and
Department of Physics and Astronomy, University of Calgary,
Calgary T2N 1N4, Alberta, Canada}
\author{Christoph Simon}
\address{Institute for Quantum Information Science and
Department of Physics and Astronomy, University of Calgary,
Calgary T2N 1N4, Alberta, Canada}

\begin{abstract}
Bohmian mechanics reproduces all statistical predictions of quantum mechanics, which ensures that entanglement cannot be used for superluminal signaling. However, individual Bohmian particles can experience superluminal influences. We propose to illustrate this point using a double double-slit setup with path-entangled photons. The Bohmian velocity field for one of the photons can be measured using a recently demonstrated weak-measurement technique. The found velocities strongly depend on the value of a phase shift that is applied to the other photon, potentially at spacelike separation.
\end{abstract}
\maketitle

Bohmian mechanics \cite{bohm} (BM) is the most famous and most developed hidden-variable theory for quantum physics. It postulates the existence of both a quantum wave, which corresponds to the usual quantum wave function, and of particles whose motion is guided by the wave, in the spirit of De Broglie. The exact positions of these particles are the additional ``hidden'' variables compared to the usual quantum physical description.

Under the assumption that the distribution of particle positions is given by the modulus squared of the wave function, which is the equilibrium state in BM \cite{bohmequilibrium}, all statistical predictions of BM agree exactly with those of standard quantum mechanics. This means in particular that the uncertainty principle applies, such that it is impossible to precisely observe the trajectory of an individual Bohmian particle.

However, in Ref. \cite{wiseman} it was pointed out that the velocity field for an ensemble of Bohmian particles, which is related to the (multi-dimensional) gradient of the wave function, can be experimentally observed in a direct and intuitive way using the concept of weak-value measurements \cite{weak}. This proposal was recently implemented in Ref. \cite{kocsis} for a double-slit experiment with single photons.

Arguably, the essence of BM is its non-local character, which was recognized already by Bohm when proposing his theory \cite{bohm2}. For entangled quantum states, actions performed on one particle can have an instantaneous effect on the motion of another particle far away. This feature motivated Bell to study the question whether all hidden-variable theories for quantum mechanics have to be non-local. This question was of course answered in the affirmative by Bell's theorem \cite{bell}.

It should be emphasized that the superluminal influences experienced by individual Bohmian particles cannot be used for superluminal signaling, as long as the particle positions are distributed according to the modulus squared of the wave function. From the point of view of BM, the theory of relativity therefore remains valid, but only in a statistical sense \cite{bohm2,nikolic}.

In this paper we propose to demonstrate this highly non-local character of BM in an experiment using entangled photon pairs. Building on the ideas of Refs. \cite{wiseman,kocsis}, we propose to use path-entangled photons and a double-double-slit setup \cite{hz}, with variable phase shifts between the two slits on one side. We show that the velocity field (and hence the trajectory) for the particle on the other side depends on the phase shift applied to the first particle, and we discuss how this can be observed experimentally, thus allowing a striking demonstration of BM's non-locality.

In BM, for a two-particle system the velocity field for particle A is given by
\begin{equation}
{\bf v}_A ({\bf x}_A,{\bf x}_B)=\frac{{\bf j}_A({\bf x}_A,{\bf x}_B)}{|\psi({\bf x}_A,{\bf x}_B)|^2},
\end{equation}
where
\begin{equation}
{\bf j}_A({\bf x}_A,{\bf x}_B)=i \psi^*({\bf x}_A,{\bf x}_B) \nabla_A \psi({\bf x}_A,{\bf x}_B) + c.c.,
\end{equation}
and $\psi({\bf x}_A,{\bf x}_B)$ is the two-particle wave function.
To obtain the velocity field for particle B, the gradient is taken with respect to the position of that particle. The velocity field ${\bf v}_A ({\bf x}_A,{\bf x}_B)$ is interpreted as giving the velocity for a Bohmian particle A at position ${\bf x}_A$, provided that particle B is at position ${\bf x}_B$. It is easy to see that the dependence on particle B's position disappears for unentangled (product) quantum states.

It is tempting to interpret the fact that for entangled quantum systems the velocity for particle A depends on the position of particle B as an immediate demonstration of the non-locality of BM. However, this is in fact not conclusive. BM is deterministic. This means that without external intervention the positions of the particles at all times are uniquely determined by their initial positions plus the initial wave function. The apparent non-locality could therefore be seen as simply an unusual form of expressing the dependence on the initial conditions. This is particularly relevant in the typical case where the particles originate from the same source, and were thus not far apart at all times. It is conceptually clearer to introduce an external local influence on one particle and study its effect on the other particle. This is the approach that we propose to pursue below.

We now explain how the velocity field can be measured. In Ref. \cite{wiseman} it was pointed out that
\begin{equation}
{\bf v}_A({\bf x}_A,{\bf x}_B)=
\frac{1}{m} \mbox{Re} \frac{\langle {\bf x}_A, {\bf x}_B|\hat{{\bf p}}_A|\psi\rangle}{\langle {\bf x}_A, {\bf x}_B|\psi\rangle}, \label{vp}
\end{equation}
where $\hat{{\bf p}}_A$ is the momentum operator for particle A, and $|\psi\rangle$ is the two-particle quantum state, such that $\langle {\bf x}_A, {\bf x}_B|\psi\rangle=\psi({\bf x}_A, {\bf x}_B)$. The analogous relation holds for ${\bf v}_B({\bf x}_A,{\bf x}_B)$ and $\hat{{\bf p}}_B$.

\begin{figure*}[h!]
\centering{}\includegraphics[width=\textwidth]{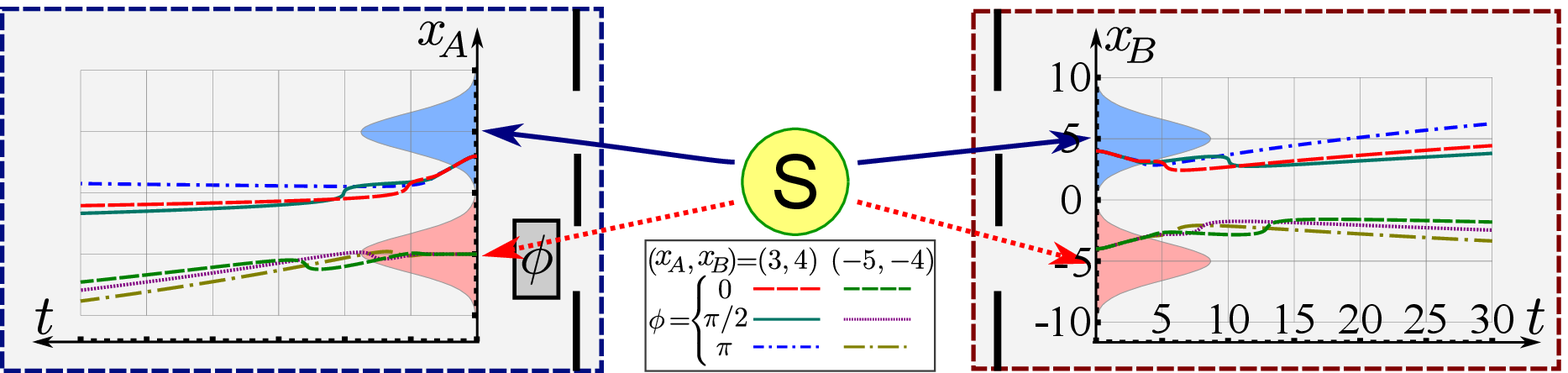}\caption{Conceptual structure of the proposed experiment and calculated Bohmian trajectories. A source produces a pair of path-entangled photons such that either both photons go through the upper slits on each side or both photons go through the lower slits, see Eq. (\ref{pathent}). A phase shift $\phi$ is introduced behind the lower slit on the left, creating the state of Eq. (\ref{phistate}). The resulting Bohmian trajectories for photons $A$ and $B$ are shown for two possible starting positions, corresponding to the Bohmian particles passing through the upper or lower slits respectively, and for different values of the phase shift. One can see that the trajectory of photon $B$ depends on $\phi$, even though the phase shift is applied to photon $A$. This shows the highly non-local character of Bohmian mechanics.} \label{conceptual}
\end{figure*}

Eq. (\ref{vp}) allows one to make a link to the weak-value formalism \cite{weak}. In this approach the system under consideration, which is initially in a given quantum state $|\psi\rangle$, is first made to interact weakly with a pointer with an interaction Hamiltonian of the general form $H=\chi \hat{p} \hat{\sigma}$, where $\chi$ is the coupling strength, $\hat{p}$ is the observable of the system that is to be measured weakly, and $\hat{\sigma}$ is an operator of the pointer. Then one performs a projective measurement of the system in some basis $\{ |\phi_k\rangle \}$. One can show that for sufficiently weak interactions the operation performed on the pointer conditional on finding a final state $|\phi_k\rangle$ of the system is then
of the form $e^{i \chi t p_w^{(k)} \hat{\sigma}}$, where $t$ is the interaction time, and the {\it weak value} $p_w^{(k)}$ is given by
\begin{equation}
p_w^{(k)}=\frac{\langle \phi_k | \hat{p} |\psi\rangle}{\langle \phi_k | \psi\rangle}. \label{pw}
\end{equation}
Comparing Eqs. (\ref{vp}) and (\ref{pw}) one sees that by identifying the system observable $\hat{p}$ with the momentum operator $\hat{{\bf p}}_A$ and the final measurement basis $\{ |\phi_k\rangle \}$ with the two-particle position basis $\{ | {\bf x}_A, {\bf x}_B\rangle \}$, the velocity field is given by the real part of the weak values of the momentum operator. In the related single-particle experiment of Ref. \cite{kocsis} the pointer was implemented by the polarization degree of freedom of the individual photons, and the weak value of the momentum was inferred from the rotation of the polarization, see below.

We will first describe the proposed experiment in conceptual terms, then we will discuss its implementation in more detail. We consider a source of pairs of entangled particles (see Figure 1). Particle A is emitted towards the left, and particle B towards the right. Each particle encounters a double slit. The source is constructed in such a way that at the time when each particle is in the plane of its respective double slit the wave function of the total system is
\begin{equation}
\frac{1}{\sqrt{2}} \left( f_u(x_A) f_u(x_B)+f_d(x_A) f_d(x_B) \right).
\label{pathent}
\end{equation}
Here we are only considering a single coordinate for each particle ($x_A$ and $x_B$ respectively), along a line connecting the two slits (transverse to the direction of motion);  $f_u(x_A)$ is the wave function corresponding to the upper slit for particle A. It has zero overlap with $f_d(x_A)$, which corresponds to the lower slit, and analogously for particle B. We will also immediately introduce a phase shifter that is placed just behind the lower slit for particle A. It causes a variable phase shift $\phi$, leading to a modified wave function
\begin{equation}
\psi_0(x_A,x_B)=\frac{1}{\sqrt{2}} \left( f_u(x_A) f_u(x_B)+e^{i\phi} f_d(x_A) f_d(x_B) \right).
\label{phistate}
\end{equation}
In the following we will study how changing $\phi$, which in our proposed experiment is done on the left, affects the Bohmian trajectories for the particle on the right.

\begin{figure}[t]
\centering{}\includegraphics[width=\columnwidth]{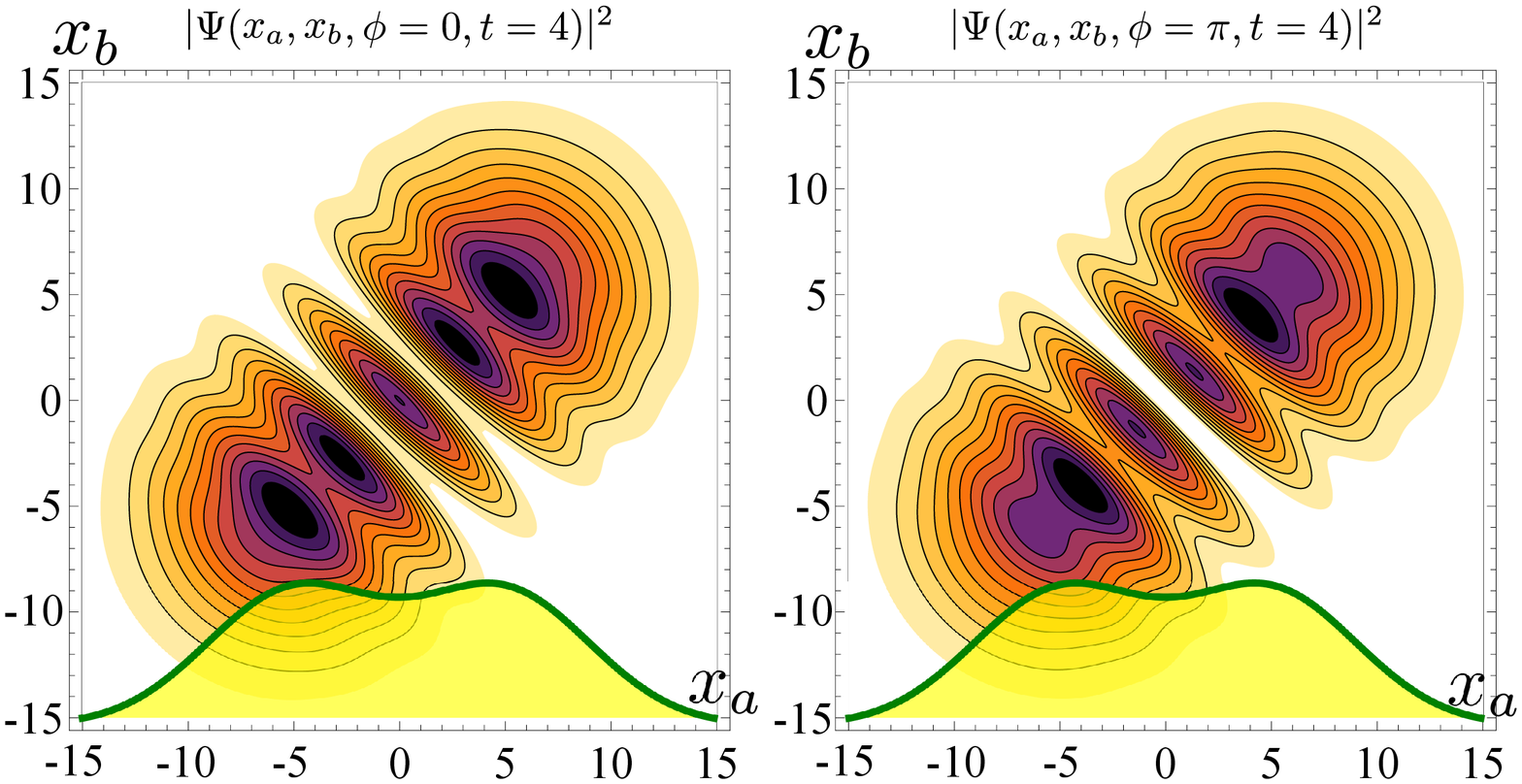}\caption{Two-particle detection probability for the experiment of Figure 1. The case $\phi=0$ is shown on the left, $\phi=\pi$ on the right. One sees interference fringes whose location clearly depends on the value of $\phi$. However, the corresponding single-particle probability distributions do not depend on $\phi$, as can be seen from the marginals shown at the bottom of the figures. This ensures that superluminal communication is impossible, see also the text.}
\label{marg}
\end{figure}

We consider a free time evolution, corresponding to a Hamiltonian $H=p_A^2/2m+p_B^2/2m$ for the considered coordinates, with Eq. (\ref{phistate}) as the initial state, where $p_A$ and $p_B$ are the momenta conjugate to $x_A$ and $x_B$. For photons, such an evolution arises naturally in the paraxial approximation \cite{garrison}, provided that one defines the ``time'' $t$ and the ``mass'' $m$ such that they satisfy $\frac{\hbar t}{m}=\frac{z}{k_0}$, where $z$ is the position in longitudinal direction and $k_0$ is the (central) longitudinal wave vector for the photons. As time evolves, or equivalently as the particles propagate away from the respective planes of the double slits, the wave packets spread (diffract) and begin to overlap. This leads to the appearance of interference fringes in the joint two-particle detection probability $p(x_A,x_B)=|\psi(x_A,x_B,t)|^2$, see Figure 2, where $\psi(x_A,x_B,t)$ is the wave function at time $t$. The exact location of the fringes depends on the phase $\phi$. The system thus exhibits two-particle interference.

In contrast, there is no single-particle interference \cite{ghz}. That is, there are no interference fringes and no dependence on $\phi$ in the marginal single-particle probability distributions $p(x_A)=\int dx_B p(x_A,x_B)$ and $p(x_B)=\int dx_A p(x_A,x_B)$, see Figure 2. Note that if $p(x_B)$, which is locally measurable, depended on the phase shift $\phi$ implemented as described above, this would in principle allow superluminal communication between the experimenter on the left and the experimenter on the right. Moreover the phase $\phi$ in Eq. (\ref{phistate}) could also be caused by a phase shifter on the right hand side. As a consequence, a dependence of $p(x_A)$ on $\phi$ would also correspond to superluminal signaling.

As discussed in the introduction, quantum physics does not allow superluminal signaling between observers, and this is true for BM as well, provided that the particles are distributed according to the square of the wave function. However, in BM there are superluminal effects at the level of the individual particles. This can be seen by studying the BM velocity field. As discussed above, one has $v_A(x_A,x_B,t)=j_A(x_A,x_B,t)/|\psi(x_A,x_B,t)|^2$, with $j_A(x_A,x_B,t)=i \psi^*(x_A,x_B,t) \frac{\partial}{\partial x_A} \psi(x_A,x_B,t) +c.c.$, and analogously for $v_B$. The key point is that, since the state is entangled, both $v_A$ and $v_B$ depend on the phase shift $\phi$, even though it can be applied in a purely local way, e.g. on the left hand side of the setup. As a consequence, for given initial Bohmian positions of particles A and B, the Bohmian trajectories diverge {\it for both particles} for different values of $\phi$. This is shown explicitly in Figure 1. Note that the double slits on the left and right can be arbitrarily far apart. As a consequence, there can be a spacelike separation between the application of the phase shift on the left and the divergence of the trajectories on the right. This clearly shows the highly non-local character of BM.

\begin{figure*}[t]
\centering{}\includegraphics[width=\textwidth]{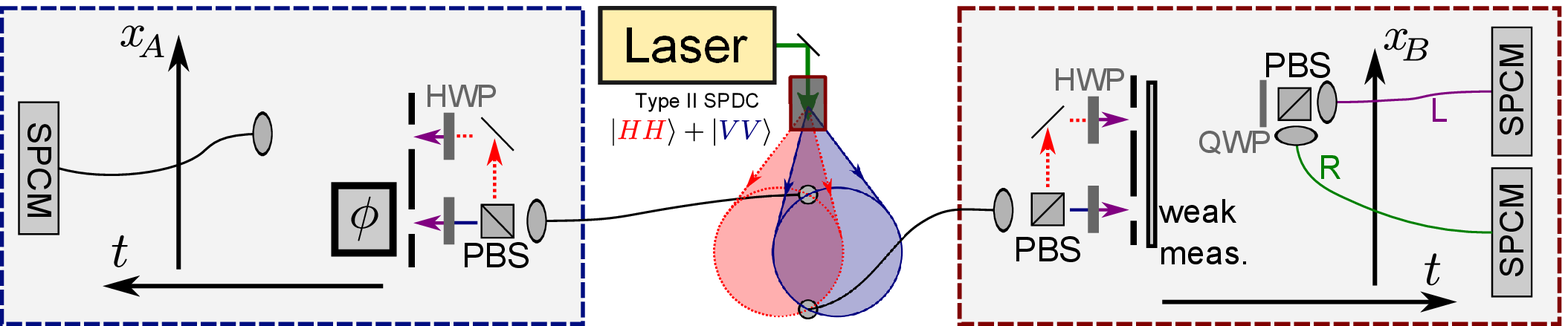}\caption{Proposed experimental realization of the conceptual setup shown in Figure 1, including a weak measurement to determine the velocity field for particle $B$. Type-II spontaneous parametric down-conversion is used to produce pairs of polarization-entangled photons, which are mode filtered using single-mode fibers. The polarization entanglement is converted to path entanglement using polarizing beam splitters (PBS) and half-wave plates (HWP). Together with the phase shifter behind one of the slits for particle $A$ this creates the state of Eq. (\ref{phistate}). The Bohmian velocity field of particle $B$ is measured by performing a weak measurement of its momentum, followed by a joint strong measurement of both particles' positions. The weak measurement is implemented using the birefringence in a calcite crystal to couple the photon's momentum to its polarization, see the text. The position measurements are performed by (multi-mode) fiber-coupled single-photon counting modules (SPCMs). For photon $B$ a polarization beam splitter and
two SPCMs are used in order to have access to the polarization information that is necessary for the weak measurement.} \label{implementation}
\end{figure*}

We now describe the proposed experiment in more detail,
as shown in Figure \ref{implementation}. By using type-II
spontaneous parametric down-conversion, one
can generate polarization-entangled pairs of photons in (ideally)
a state $\frac{1}{\sqrt{2}}(|H\rangle |H\rangle+|V\rangle |V\rangle)$ \cite{kwiat,fedrizzi}. The two photons are then
coupled into single-mode fibres, in order to eliminate any potentially existing correlations between the spatial structure of the photon
wavepackets and their polarization states. After the fibres,
the use of polarizing beamsplitters and half-wave
plates allows this state to be converted into the path-entangled state of Eq. (\ref{pathent}), with all the information transferred from the
polarization states into spatial states corresponding to the upper and lower slit on each side. A phase shifter placed behind the lower slit for photon $A$ then creates the state of Eq. (\ref{phistate}).

As the photons are no longer entangled in polarization,
the polarization degree of freedom of photon $B$ is available for measuring the Bohmian velocity field in a manner analogous
to Ref. \cite{kocsis}. The key element is a piece of birefringent material such as calcite. As explained in detail in Ref. \cite{kocsis}, the calcite causes a polarization rotation of the photon that is proportional to the weak value of the transverse momentum. Since photons $A$ and $B$ are entangled in their external variables, the weak value, and hence the Bohmian velocity field, depends on the measured position values for both photons, see Eqs. (\ref{vp}) and (\ref{pw}). Photon $B$ has to be detected in a polarization sensitive manner in order to determine the angle by which its polarization was rotated, and thus the value of the Bohmian velocity.

By moving the photon detectors in longitudinal and transverse direction, one can map the velocity field. For a fixed $\phi$ we estimate that one would need 25 different longitudinal positions (which can be varied jointly on each side since the photons have the same longitudinal velocity), 40 bins in each plane, and 1000 detected photon pairs for each combination of bins, to get a complete picture of the trajectories similar to what was done in Ref. \cite{kocsis}. Using a highly efficient photon pair source such as the one in Ref. \cite{fedrizzi} with of order 10 mW pump power, one can in principle achieve detected coincidence rates of 1 MHz. This would be reduced by a factor of 1600 for the above number of bins per plane. Collecting a sufficient number of detected pairs for all planes and bin combinations would then take about $25\times 40^2\times 1000 \times \frac{40^2}{10^6}$ seconds, i.e. about 18 hours.

\begin{figure}[t]
\centering{}\includegraphics[width=\columnwidth]{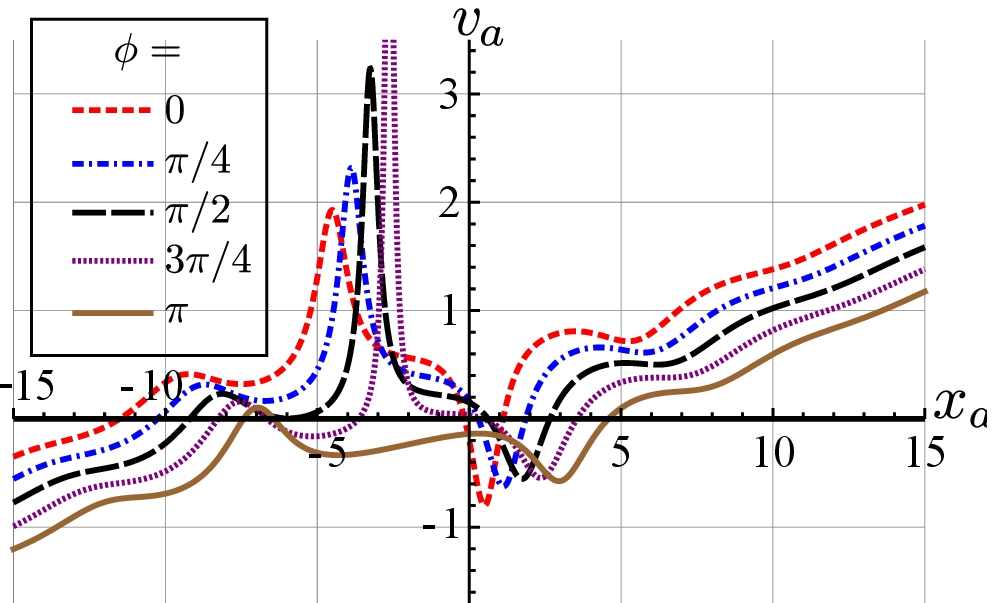}\caption{Velocity profiles that could be measured using the apparatus in Figure 3. For a fixed position of the detector on the $A$ side ($t=4, x_A=4$), the (Bohmian) velocity field for photon $B$ is measured at different transverse positions $x_B$. One sees that the velocity depends strongly on the phase $\phi$, even though the phase is applied on the $A$ side. The velocity profiles are offset from each other vertically by 0.2 for clarity, with the lowest ($\phi=\pi$) profile being at its true position.} \label{vels}
\end{figure}

Another option that is less demanding in terms of detected pairs is to just measure the velocity profile in a fixed plane. This would also be sufficient to show the effect of changing $\phi$ on the motion of a remote Bohmian particle. Fig. \ref{vels} shows the theoretical predictions for such a scenario. 

For either approach, spacelike separation between the two sides could be implemented by using long fibers before the double slits and a fast electro-optic phase modulator in analogy with Ref. \cite{weihs}.

We have proposed a clear and feasible experimental demonstration of the highly non-local character of Bohmian mechanics. We have discussed an experiment based on photons, but an implementation with atoms may be possible as well based on a source of the type discussed in Ref. \cite{kofler}. It should be emphasized that the expected results are fully compatible with standard quantum physics. Our proposal thus has no direct bearing on the question whether the Bohmian trajectories correspond to something ``real'', or more generally whether there indeed are hidden variables underlying quantum physical phenomena. However, in the class of hidden-variable theories compatible with quantum physics, Bohmian mechanics stands out by being both historically influential and highly developed. Its position is further strengthened not only by Bell's theorem, which showed that all such theories have to be non-local, but also by two much more recent results. It is a consequence of the work of Ref. \cite{colbeck} that all non-trivial hidden-variable theories compatible with quantum physics have to include superluminal influences at the level of the hidden variables (they cannot be ``no-signaling''). Furthermore Ref. \cite{PBR} showed that every such hidden-variable theory has to include the wave function as a variable. Several features of Bohmian mechanics that might have seemed unattractive at first sight have thus been shown to be unavoidable -- if a hidden-variable description of nature is desired.

We thank A. Aspect, R. Colbeck, D. Mahler, S. Popescu, R. Renner, L. Rozema, A. Steinberg, H. Wiseman, and A. Zeilinger for useful discussions. This work was supported by AITF and NSERC. B.B. would like to thank Tom Frank for supporting his fellowship during which this work was done.

%%%%%%%%%%%%%%%%%%%%%%%All the bibliography is in MM.bib in the same folder%%%%%%%%%%%%%%%%%%%%%%%

%%%%%%%%%%%%%%%%%%%%%%%All the bibliography is in MM.bib in the same folder%%%%%%%%%%%%%%%%%%%%%%%

\end{document}